\documentclass[12pt]{article}
\usepackage{epsfig,rotate}
\let\section=\subsection
\let\subsection=\subsubsection

\begin{document}
\begin{center}
{\large \bf STRANGE QUARK MATTER WITH}\\[2mm]
{\large \bf EFFECTIVE QUARK MASSES\footnote{Supported by BMBF, GSI Darmstadt, and DFG}}\\[5mm]
K.~SCHERTLER, C.~GREINER, and M.~H.~THOMA{\footnote{Heisenberg Fellow}} \\[5mm]
{\small \it Institut f\"ur Theoretische Physik, Universit\"at Giessen \\
35392 Giessen, Germany \\[8mm] }
\end{center}

\begin{abstract}
\noindent
The properties of strange quark matter at zero temperature are investigated 
including medium effects. The quarks are considered as quasiparticles which acquire an 
effective mass generated by the interaction with the other quarks of the 
dense system. Within this approach we find that these medium effects reduce 
the binding energy of strange quark matter with respect to $^{56}F\!e$.  
\end{abstract}

\section{Introduction}
Strange quark matter (SQM) has been suggested as a possible stable or meta\-stable 
phase of nuclear matter \cite{SQM}, which might be realized in form  of small
droplets (stranglets) or in the interior of superdense stars \cite{MH}.
Applying the Fermi gas model for quarks, the energy per baryon of such a
system might be lower than the one of $^{56}F\!e$ due to the conversion of
up and down quarks into strange quarks by the weak interaction. The
equation of state of this system has been described as a non-interacting 
gas of quarks at zero temperature, taking into account the bag constant \cite{SQM,FJ}.
Also quark interactions within lowest order perturbative QCD have been 
considered \cite{FJ,FM}.

In condensed matter as well as in nuclear physics medium effects play an 
important role. One of the most important medium effects are effective
masses generated by the interaction of the particles with the system.
In a gluon gas at high temperature the consideration of an effective
mass for the gluons within the ideal gas approximation
leads to an excellent description of the equation of state found in lattice 
calculations \cite{GS,PKPS}.

Here we want to apply the idea of an ideal gas of quasiparticles with
effective masses to the case of quark matter. Instead of gluons at
finite temperature we consider up, down, and strange quarks at zero 
temperature but finite chemical potential $\mu $. 

\section{Effective Quark Masses}
The aim of the present section is the derivation of an effective quark mass
in an extended, dense system of quarks at zero temperature. This effective 
mass is defined as the zero momentum limit of the quark dispersion relation
following from the poles of the effective quark propagator
\begin{equation} \label{prop}
S^*(p)=\frac{1}{p_\mu \gamma ^\mu -m-\Sigma},
\end{equation}
where $m$ is the current quark mass and $\Sigma $ the quark self energy.
The self energy is calculated within the hard dense loop (HDL)
approximation \cite{MA}, where only hard momenta of the order of the quark
chemical potential $\mu $ for the internal quark and gluon lines in the 
one-loop quark self energy are used. As in the high temperature case 
\cite{BP} the vacuum contribution and the contribution from integration over
soft momenta are of higher order. This approximation is equivalent to the
leading term in the high density approximation analogously to the finite 
temperature case \cite{KW}. In this way we arrive at a gauge invariant 
expression for the self energy, complete to leading order in the coupling
constant $g$. Of course, due to the perturbative nature of this approximation,
the resulting  expressions are only valid for small coupling constants,
corresponding to a large chemical potential $\mu $, i.e. high density,
according to asymptotic freedom. However, as shown in the case of a hot
gluon gas, the equation of state using effective masses calculated
perturbatively agrees very well with lattice results even at temperatures
where the coupling constant is not small \cite{PKPS}. 

In the case of a vanishing current quark mass, as it can be assumed for
up and down quarks, the effective mass following from the HDL quark self energy 
is given by \cite{KW,VT}
\begin{equation} \label{defmq}
{m_q^*}^2=\frac{g^2\mu ^2}{6\pi ^2}.
\end{equation}

For strange quarks we adopt an effective
quark mass following from the dispersion relation of massive quarks, where we used the HDL
self energy assuming the current quark mass to be soft ($m_s \sim g\mu $) \cite{BO}:
\begin{equation} \label{defms}
m_s^*=\frac{m_s}{2}+\sqrt {\frac{m_s^2}{4}+\frac{g^2\mu ^2}{6\pi ^2}}.
\end{equation}

\section{Equation of State}

At temperature $T=0$ the particle density $\rho$ and energy density $\epsilon$ 
of a Fermi gas are given by \cite{KAP}

\begin{equation} \label{defrhogeneral}
\rho(\mu)= \frac{d}{6 \pi^2} k_F^3, 
\end{equation}

\begin{equation} \label{defepsgeneral}
\epsilon(\mu) - B^*(\mu)=\frac{d}{16 \pi^2} \left[ \mu \, k_F \left( 2 \mu^2 - {m^*}^2 \right)
   - {m^*}^4 \ln {\left( \frac{k_F + \mu}{m^*}\right)} \right] 
\end{equation}


Here $d$ denotes the degree of degeneracy (e.g. $d=6 n_f$ for $n_f$ flavors). Up to the 
additional function $B^*$, which can be regarded as a $\mu $-dependent bag constant,
these are the
ideal Fermi gas formulas at temperature $T=0$ for quasiparticles of mass
$m^*$ and chemical potential $\mu$. The Fermi momentum is $k_F = (\mu^2 -
{m^*}^2)^{1/2}$. Due to the $\mu$-dependence of $m^*(\mu)$ the functions $B^*(\mu)$
are necessary to maintain thermodynamic self-consistency
\cite{GorensteinYang1995}. 

In the case of $m^*(\mu)= m^*_q(\mu)$ given by (\ref{defmq}) we found \cite{SGT}

\begin{equation}
  B^*_q(\mu_q) = -\frac{d_q}{16 \pi^2} \left[ \alpha^2 \beta - \alpha^4 \ln 
  {\left(\frac{\beta+1}{\alpha} \right)} \right] \mu_q^4 \, ,
\end{equation}
where $\alpha = g/(\sqrt{6} \pi)$ and $\beta =\sqrt{1-\alpha^2}$. 
The equations of state (\ref{defrhogeneral})-(\ref{defepsgeneral}) then assume the simple form

\begin{equation} \label{defrhoq}
  \rho_q(\mu_q) = \frac{d_q}{6 \pi^2} \beta^3 \mu_q^3
\end{equation}

\begin{equation} \label{defepsq}
  \epsilon_q(\mu_q) = \frac{d_q}{8 \pi^2} \beta^3 \mu_q^4
\end{equation}


They are the well known equations for massless fermions up to the
factor $\beta^3$ (\mbox{$0<\beta(g)<1$}). Note that $\rho_q$ and $\epsilon_q$ 
decrease with increasing coupling constant $g$. 

In the case of an effective strange quark mass the additional function  $B_s^*(\mu _s)$
assumes a more complicated form given in ref.\cite{SGT}.

\section{Strange Quark Matter with Effective Masses}

Now we use the equations of state derived in the last section to calculate the energy per 
baryon of strange quark matter assuming a finite s-quark mass:

\begin{equation} \label{EdurchAs}
  \left(\frac{E}{A}\right)(\mu_q, \mu_s) = \frac{\epsilon_q(\mu_q) +\epsilon_s(\mu_s)  +
    B_0}{\rho_B(\mu_q, \mu_s)}.
\end{equation}
Here $B_0$ denotes the bag constant.
The baryon density is now given by $\rho_B=(\rho_q(\mu_q)+\rho_s(\mu_s))/3$.
As usual \cite{Greiner} we define the strangeness fraction as 
$f_s=\rho_s(\mu_s)/\rho_B(\mu_q,\mu_s)$.
Fig.1 shows the minimum energy per baryon (corresponding to a vanishing pressure)
over $f_s$ for various values of $g$. 
The bag constant is assumed to be $B_0^{1/4}=145$ MeV and the
current s-quark mass as $m_s=150$ MeV. One
sees that $E/A$ increases with increasing $g$. Consequently, the absolute minimum of the 
energy (Fig.2), obtained 
from the optimal $f_s$, increases for example from 874 MeV ($g=0$) to 943 MeV 
($g=3$). 
Therefore, for the chosen parameters and realistic values of the coupling constant
SQM is not bound with respect to $^{56}Fe$. For larger $m_s$ and $B_0^{1/4}$ even 
larger values for $E/A$ are obtained.
Furthermore, the baryon density is found to decrease with 
increasing $g$ while the energy density $\epsilon$ stays approximately 
constant \cite{SGT}. The strangeness fraction $f_s$ decreases with 
increasing $g$. 

\begin{figure}[ht]
\centerline{\rotate[r]{\epsfig{file=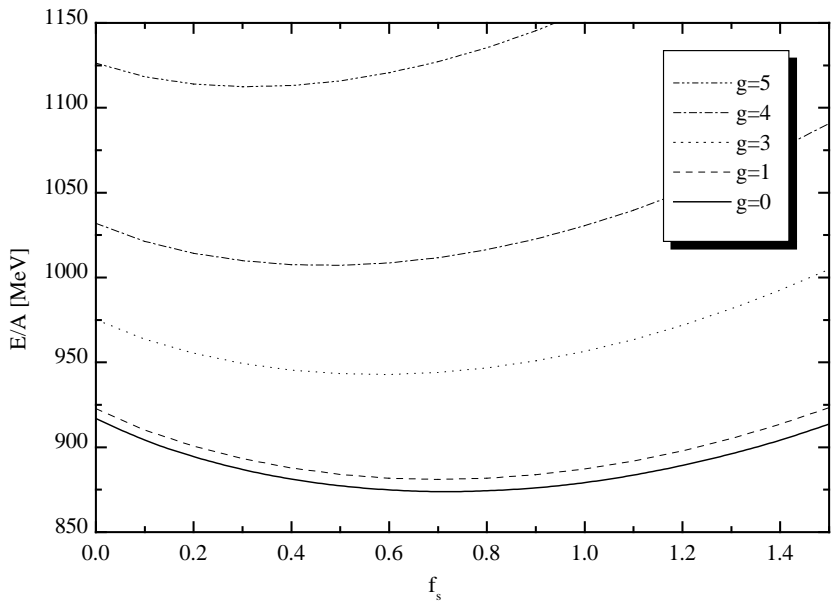,height=13cm}}}
\caption{Minimum energy per baryon over the strangeness fraction at $B_0^{1/4}=145$ MeV
and $m_s=150$ MeV}
\label{eafs}
\end{figure}

\begin{figure}[ht]
\centerline{\rotate[r]{\epsfig{file=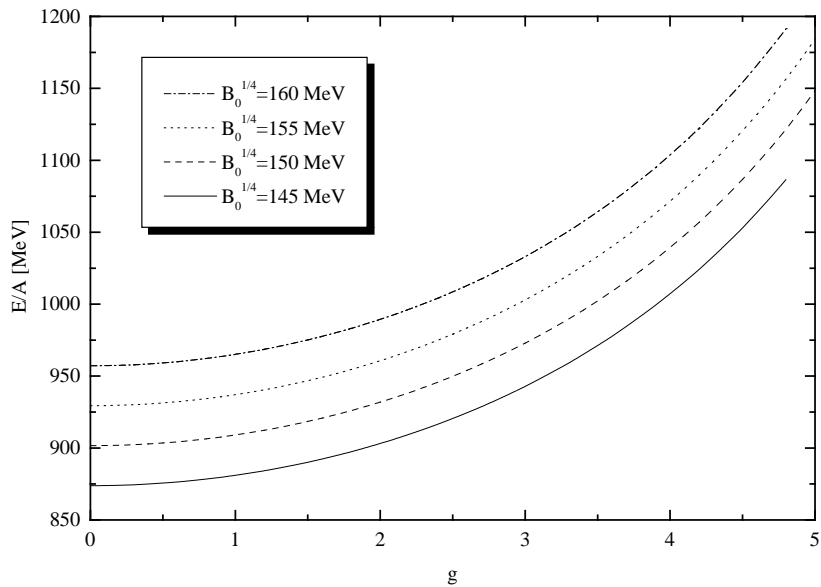,height=12.5cm}}}
\caption{Absolute minimum of energy per baryon over the coupling constant at $m_s=150$ MeV}
\label{eag}
\end{figure}

It is not possible to describe the influence of medium effects just by altering the bag 
constant \cite{SGT}. The same observation holds for a variation of $m_s$ instead of $B_0$.
Furthermore, the EOS of SQM has been computed including $\alpha_s$-corrections, i.e. one-gluon
exchange \cite{FM, Greiner}. Increasing $\alpha _s$ there leads to an increase of $E/A$ as in 
our 
approach. However, at the same time $f_s$ increases in contrast to our case. Hence, we conclude
that the influence of the effective quark masses on SQM cannot be simulated by taking
into account $\alpha _s$-corrections or changing $B_0$ or $m_s$.

Concluding, our investigations suggest that SQM is not absolute 
stable if medium effects are taken into account. However, it might still be metastable,
i.e. stable against decays caused by the strong interaction, which might have interesting
consequences for the formation of strangelets in ultrarelativistic heavy ion collisions
\cite{Greiner}. 

Finally, applying our results to strange quark stars, we observe \cite{SGT} that the 
radius-mass relation of
these stars is hardly changed by the presence of an effective quark mass since the 
pressure as function of the energy density, entering into the Oppenheimer-Volkoff-Tolman
equations, 
depends only very weakly on it. However, owing to the increase of $E/A$ in the entire
star a phase transition to hadronic matter will take place at a smaller
radius in the interior of the star.

\end{document}